\documentclass[aps,prb,twocolumn,showpacs,superscriptaddress,footinbib,10pt]{revtex4-1}

\usepackage{graphicx}
\usepackage{dcolumn}
\usepackage{bm}
\usepackage{amssymb}
\usepackage{amsmath}
\usepackage{natbib}
\usepackage[french]{babel}

\newcommand{\muu}{cm$^2$/(Vs)\,}

\begin{document}

\title{Angular-Resolved Photoemission Electron Spectroscopy and transport studies of the elemental topological insulator $\alpha$-Sn.}

\author{Quentin~Barbedienne}
\affiliation{Unit\'e Mixte de Physique, CNRS, Thales, Univ. Paris-Sud, Universit\'e Paris-Saclay, Palaiseau 91767, France}
\author{Julien~Varignon}
\affiliation{Unit\'e Mixte de Physique, CNRS, Thales, Univ. Paris-Sud, Universit\'e Paris-Saclay, Palaiseau 91767, France}
\author{Nicolas~Reyren}
\affiliation{Unit\'e Mixte de Physique, CNRS, Thales, Univ. Paris-Sud, Universit\'e Paris-Saclay, Palaiseau 91767, France}
\author{Alain~Marty}
\affiliation{Univ. Grenoble Alpes, CNRS, CEA, Grenoble INP, INAC-SPINTEC, F-38000 Grenoble, France}
\author{C\'eline~Vergnaud}
\affiliation{Univ. Grenoble Alpes, CNRS, CEA, Grenoble INP, INAC-SPINTEC, F-38000 Grenoble, France}
\author{Matthieu~Jamet}
\affiliation{Univ. Grenoble Alpes, CNRS, CEA, Grenoble INP, INAC-SPINTEC, F-38000 Grenoble, France}
\author{Carmen~Gomez-Carbonell}
\affiliation{Centre de Nanosciences et de Nanotechnologies,
CNRS, Univ. Paris-Sud, Universit\'e Paris-Saclay,
C2N-Marcoussis, Route de Nozay, 91460 Marcoussis, France}
\author{Aristide~Lema\^{i}tre}
\affiliation{Centre de Nanosciences et de Nanotechnologies,
CNRS, Univ. Paris-Sud, Universit\'e Paris-Saclay,
C2N-Marcoussis, Route de Nozay, 91460 Marcoussis, France}
\author{Patrick~Le~F\`evre}
\affiliation{Synchrotron SOLEIL, L'Orme des Merisiers, 91192, Gif-sur-Yvette, France}
\author{Fran\c{c}ois~Bertran}
\affiliation{Synchrotron SOLEIL, L'Orme des Merisiers, 91192, Gif-sur-Yvette, France}
\author{Amina~Taleb-Ibrahimi}
\affiliation{Synchrotron SOLEIL, L'Orme des Merisiers, 91192, Gif-sur-Yvette, France}
\author{Henri~Jaffr\`es}
\affiliation{Unit\'e Mixte de Physique, CNRS, Thales, Univ. Paris-Sud, Universit\'e Paris-Saclay, Palaiseau 91767, France}
\author{Jean-Marie~George}
\email{jeanmarie.george@cnrs-thales.fr}
\affiliation{Unit\'e Mixte de Physique, CNRS, Thales, Univ. Paris-Sud, Universit\'e Paris-Saclay, Palaiseau 91767, France}
\author{Albert~Fert}
\affiliation{Unit\'e Mixte de Physique, CNRS, Thales, Univ. Paris-Sud, Universit\'e Paris-Saclay, Palaiseau 91767, France}
\date{\today}

\begin{abstract}
Gray tin, also known as $\alpha$-Sn, can be turned into a three-dimensional topological insulator (3D-TI) by strain and finite size effects. Such room temperature 3D-TI is peculiarly interesting for spintronics due to the spin-momentum locking along the Dirac cone (linear dispersion) of the surface states. Angle resolved photoemission spectroscopy (ARPES) has been used to investigate the dispersion close to the Fermi level in thin (0\,0\,1)-oriented
epitaxially strained films of $\alpha$-Sn, for different film thicknesses as well as for different capping layers (Al, AlO$_x$ and MgO). Indeed a proper capping layer is necessary to be able to use $\alpha$-Sn surface states for spintronic applications. In contrast with free surfaces or surfaces coated with Ag, coating the $\alpha$-Sn surface with Al or AlO$_x$ leads to a drop in the Fermi level below the Dirac point, an important consequence for electronic transport is the presence of bulk states at the Fermi level. $\alpha$-Sn films coated by AlO$_x$ are studied by electrical magnetotransport: despite magnetotransport properties of the bulk electronic states of the $\Gamma _8$ band playing an important role, as suggested by {\it ab-initio} calculations, there is clear evidence of surface states revealed by Shubnikov-de~Haas oscillations corresponding to ARPES observation.
\end{abstract}

\keywords{Topological Insulator; Angle Resolved Photoemission Spectroscopy; Magneto-transport measurement; ab-initio calculation}

\pacs{03.65.Vf , 79.60.Bm , 31.15.E- , 71.18.+y}
\maketitle

Classical spintronics generally uses magnetic materials to produce a
spin current from a flow of charges in the same direction. Nowadays, spin-orbit coupling (SOC) provides new directions to
generate pure spin currents~\cite{Dyakonov71,Hirsch99,Valenzuela06} in
the transverse direction. Indeed the SOC, a
relativistic correction to the equations of quantum mechanics, can be
significantly large in materials containing heavy atoms. It turns out that an efficient spin-charge conversion can be obtained
by exploiting the SOC-induced properties of two-dimensional electron
systems (2DES) found at some peculiar surfaces and interfaces. In
particular, the so-called Rashba interfaces as well as the surfaces of 3D-TI~\cite{Kane05,Konig07}
are of great interest.

During the last decade TIs have been widely studied for their
unconventional electronic properties leading to the discovery of
quantum anomalous Hall effect~\cite{Bernevig06}. Among the predicted
TIs, the 3D-TI were actively investigated since
they are, to some extent, easier to
fabricate~\cite{Chen09,Zhang09}. They become insulating because of the
opening of the gap often attributed to the strain\citep{Barfuss13}, however the band
inversion generally attributed to SOC leads to the existence of a
conducting surface states with a linear dispersion forming the
so-called Dirac Cone (DC). On the other hand, the combination of the broken inversion
symmetry and the existence of a strong electric field perpendicular to
the interface results in a peculiar connection between the spin
momentum $\sigma$ and the $\textbf{k}$ momentum vector with the time reversal symmetry imposing $E(\bf{k},\bf{\sigma})$ $=E(-\bf{k},-\bf{\sigma})$. This is sometimes called the spin-momentum locking and usually
$\bf{\sigma}\perp \bf{k}$. \cite{Qi11} Examples of prototypical
materials currently studied are Bi based materials. Bi surface states
with linear dispersion have been clearly observed using Angle Resolved
PhotoEmission Spectroscopy (ARPES)
techniques~\cite{Hsieh09}. However, magnetotransport experiments
revealed the difficulties to obtain such materials keeping the bulk state as a small enough spurious contribution to allow the surface
states properties to be clearly isolated~\cite{Analytis10}. Recently,
it was demonstrated with success, that surface states of $\rm{BiSe}_{\rm{x}}\rm{Te}_{1-\rm{x}}$
can be exploited to efficiently convert the charge flow into a
sizable spin current~\cite{Kondou10}. Moreover, in the case of materials, such as BiSeTe,
the produced spin accumulation can reverse the magnetization of an
adjacent ferromagnetic layer~\cite{Wang17,Han17}. Currently, many
questions are still open regarding the exact origins and mechanisms of
the spin-charge conversion process, the role of interfaces and in
particular how an encapsulating layer in contact with a TI might
modify the properties of the surface state itself.

In this article, we report on the study of epitaxial $\alpha$-Sn, where surface states with a characteristic spin
momentum locking have been already reported
previously~\cite{Ohtsubo13,RojasSanchez16,Barfuss13,Scholz}. In
particular, we studied by ARPES technique the impact of a capping layer on the surface states, using either insulating (AlO$_x$ or MgO), or metallic (Al or Ag) thin films. We demonstrate that the growth of one single atomic layer (1\,ML) of Al leads
to a drop in the Fermi level below the Dirac point (DP), where bulk
$p$-states coexist with the Dirac states. We also demonstrate, despite part of the electrical transport properties being governed by the bulk states, the strong evidence of the presence of the surface states by ARPES and the corresponding Shubnikov-de~Haas oscillations. From {\it ab-inito} calculations, we attribute those bulk states to the $p$-type $\Gamma _8$ multiband sheets in relatively close agreement with Ref.~\onlinecite{Barfuss13}.

We divided the article into four different sections. In the first section, we describe the preparation of samples, made at the CASSIOPEE ARPES beamline (Soleil synchrotron, France). In section II, we discuss the results obtained from \textit{in-situ} ARPES measurements. In section III, we report on the electrical transport properties of an AlO$_{x}$ coated sample (Shubnikov-de~Haas oscillations, magnetoresistance, and the Hall effect). In the last part we present an \textit{ab-initio} investigation of an $\alpha$-Sn surface state using multilayer slab technique calculations.

\section{Samples preparation}

The standard substrate used in previous reports~\cite{Barfuss13,Ohtsubo13} to stabilize the $\alpha$-phase of Sn is InSb, characterized by a small gap of about 175 meV~\cite{Kane57}. However InSb contributes to the electronic transport by shunting a part of the current of the surface state from room temperature down to 4\,K. In order to probe the magnetotransport properties of the 2DES and hence to minimize the shunting effect of the substrate, we have then fabricated (0\,0\,1)-oriented InSb thin film on a GaAs insulating substrate by molecular beam epitaxy (MBE) growth method. The chosen InSb film thickness is 120\,nm corresponding approximately to the thickness needed to recover the in-plane lattice parameter of bulk InSb ($a=6.48$\,\AA{}), free of emerging dislocations. Because this artificial substrate is prepared in another laboratory (by MBE), an amorphous capping layer of As was deposited on top to prevent any surface oxidation before being transferred to the MBE chamber where $\alpha$-Sn is grown. We then prepared the InSb film surface in the same conditions as for usual InSb substrates~\cite{RojasSanchez16,Ohtsubo13}. 
Once the amorphous As encapsulation is removed, the InSb film surface is prepared in the same manner than for usual InSb substrates~\cite{Ohtsubo13,RojasSanchez16}. A $2 \times 8$ reconstructed surface is obtained after a series of successive Ar ion etching and annealing. At this stage, the crystalline surface quality is indistinguishable from the bulk InSb substrates using LEED and RHEED: no degradation of the surface can be observed during this treatment. A single Bi atomic layer is deposited on the ``substrate", foregoing the growth of $\alpha$-Sn, as suggested by Ohtsubo {\it et al.}~\cite{Ohtsubo13}. RHEED oscillations were sometimes observed during the growth of $\alpha$-Sn corresponding to a growth rate in agreement with the one derived from the quartz balance monitor. The final $\alpha$-Sn surface was characterized by clear LEED and RHEED patterns, indicating that,on the length-scale of these probes, a monocrystalline $\alpha$-Sn is obtained as expected.

\section{Angle resolved photoemission spectroscopy experiments}

We now describe ARPES spectroscopy measurements analysis of the prepared $\alpha$-Sn samples.

\subsection{$\alpha$-Sn free surface}

\begin{center}
\begin{figure}
\includegraphics[scale=0.34]{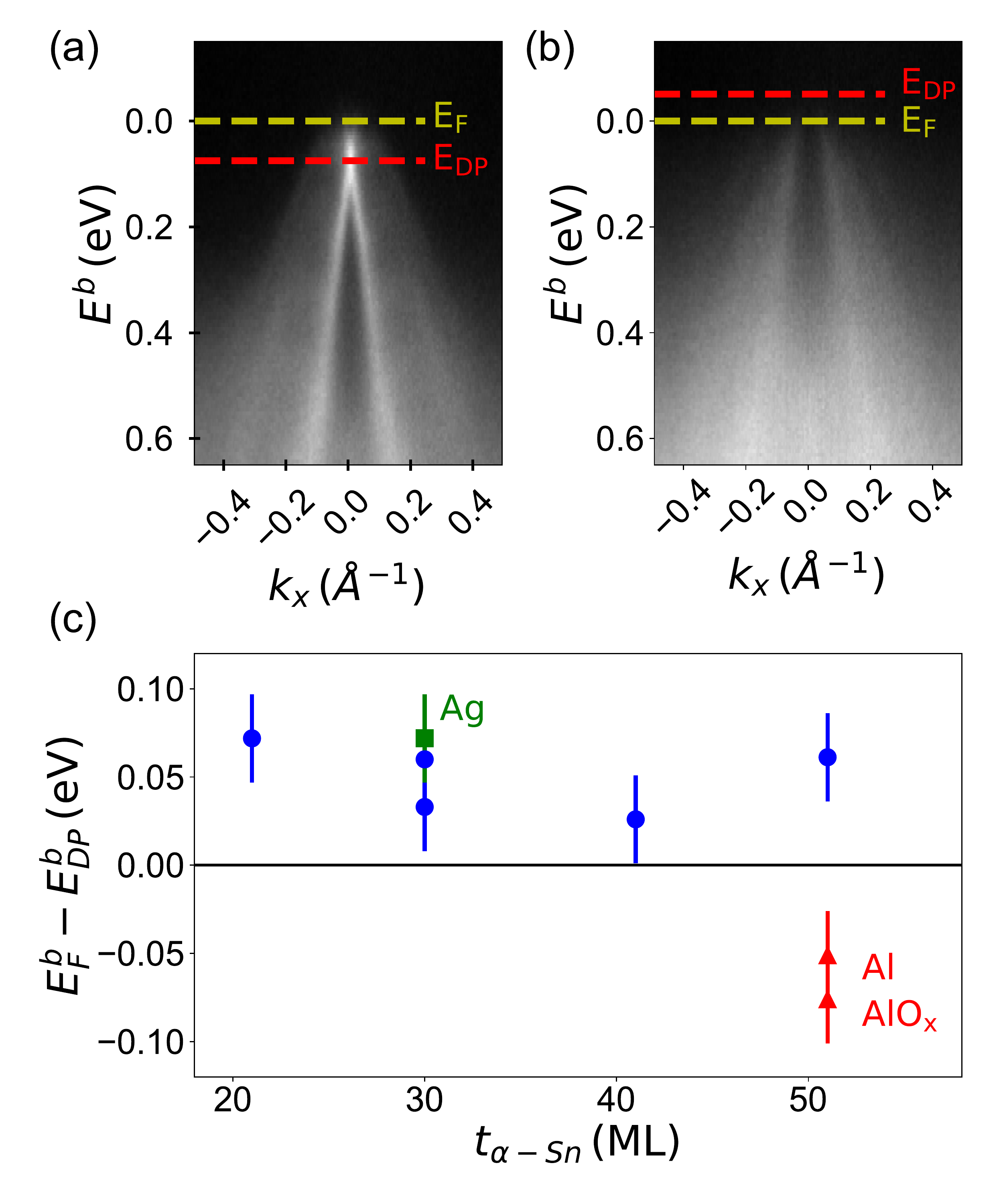}
\caption{ARPES measurement of the DC without or with capping
  layers.(a) $\alpha$-Sn(51\,ML) ARPES intensity map as a function of binding energy and momentum along the [1\,0\,0] crystalline direction (b) Corresponding map after 1\,ML Al deposition. (c) Relative position of the Fermi energy($E_{\rm F}^{\rm b}$) compared to the Dirac point energy ($E_{\rm DP}^{\rm b}$) as a function of the $\alpha$-Sn thickness (blue disks for free surface, red triangle for Al and AlO$_x$ capping, green square for Ag capping).}
\label{figure1}
\end{figure}
\end{center}

As shown on Fig.~\ref{figure1}(a) corresponding to 51 $\alpha$-Sn monolayers (ML), ARPES measurements using 19\,eV incident photons clearly reveals the existence of the Dirac cone (DC) associated with the surface state of the topological surface as already observed in previous studies~\cite{Ohtsubo13,RojasSanchez16,Barfuss13}. The DC was evidenced in all synthesized samples, characterized by ARPES, suggesting that the surface state is robust and intrinsic to the $\alpha$-Sn surface. Moreover, it also confirms the good reliability and reproducibility of the sample preparation. This validates the method and possibility to synthesize $\alpha$-Sn on a more insulating InSb(120\,nm)$|$GaAs(0.5\,mm) ``artificial substrate'' compared to bulk InSb.

From the linear dispersion visible in Fig.~\ref{figure1}(a) and (b), it becomes possible to extract the energy position of the Dirac point with respect to the Fermi level, as depicted in Fig.~\ref{figure1}(c) (blue dots). For all thicknesses investigated, ranging from 20 to 51\,ML, a 2DES was observed with a Fermi level $E_{\rm F}$ located at about, 50\,meV, within error margin, above the Dirac point energy $E_{\rm DP}$ (Fig.~\ref{figure1}(c)).

\subsection{Band structure modification by Al and AlO$_x$ capping layers}

We discuss now the observed effects of MgO, Al and AlO$_x$ capping layers on the encapsulated $\alpha$-Sn surface. We first grew 1\,ML of MgO by MBE from a stoichiometric MgO target. No surface state could then be observed by ARPES. We assign the DC disappearance to the oxidation of the top Sn layer as evidenced by Sn-3$d_{3/2}$ and Sn-3$d_{5/2}$ core levels x-ray photoemission spectroscopy (XPS) performed at the Sn $M_4$ and $M_5$ edges (see appendix). The ARPES technique cannot provide information about the potential surface states of Sn, buried under a too thick SnO$_x$ layer. In contrast, starting with the deposition of an Al layer, preserving the surface state [Fig.~\ref{figure1}(b)], we observe that the Sn oxidation is prevented, even with an extremely thin Al layer. When the Al layer is oxidized, the DC is still observed by ARPES through the thin AlO$_x$ layer.

In details, according to ARPES data and core-level spectroscopy, 1\,ML of Al is enough to prevent the oxidation of Sn. The oxidation of the 1\,ML Al can be achieved by introducing molecular O$_2$ in the MBE chamber at a pressure of $10^{-6}$\,mbar during 40 minutes. No change in the Sn photoemission spectra was observed regarding the characteristic features of the 3$d_{5/2}$ and 3$d_{3/2}$ $\alpha$-Sn core levels when covered with 1\,ML of Al or 1\,ML of oxidized Al. By contrast, the oxidation of Al can be clearly observed at the 1$s$ and 2$p_{3/2 , 5/2}$ core levels when compared to the Al metallic spectra (see the appendix).
A very important result is the observation of a Fermi level shift as visible in Fig.~\ref{figure1}(c).
In the same figure, the relative position of the Fermi energy after Ag deposition (from a previous report, Ref.~\onlinecite{RojasSanchez16}) is also plotted for comparison.

\section{Electrical transport measurements ({$\textbf{AlO}_x$} coating)}

We now focus on the electrical and magnetotransport properties of $\alpha$-Sn samples encapsulated in AlO$_x$, grown on InSb template. The magnetotransport data deal with Hall effect, Shubnikov-de~Haas oscillations, and magnetoresistance. Thanks to the encapsulation technique by the thin AlO$_x$, it becomes possible to investigate magnetotransport measurements in the samples extracted from the ARPES chamber. Two different samples have been investigated leading to similar transport properties. In particular, in this paper, we will focus on a specific sample made of AlO$_x|\alpha$-Sn (51\,ML$\approx6.55$\,nm)$|$InSb(120\,nm)$|$GaAs(substrate). We patterned Hall bars using UV optical lithography with photoresist mask and Ar ion milling with track widths of $w=200\,\mu$m and longitudinal distances between the voltage probes of $l=500\,\mu$m~. A sketch of typical device is displayed in the inset of Fig.~\ref{figure3}(a). 

In more details, the resistance measurements were performed at constant current, in the range of $10-100\,\mu$A, in a four-probe configuration. Two main features clearly appear. First, in the perpendicular geometry where a strong magnetic field $\mu_0 H_{z}$ is applied along the normal to the layers, the longitudinal resistance, $R_{xx}$ (the current flows along $x$), displays characteristic oscillations corresponding to expected Shubnikov-de~Haas (SdH) oscillations as already observed in 3D TI \cite{Qi11}. Second, equivalent oscillations are also observed and even better defined in the Hall effect geometry with the measurement of the transverse resistance $R_{xy}$ (voltage drop along $y$). The slope of $R_{xy}(H_z)$ around zero field clearly indicates a $p$-type conduction. However, from those raw data, one may extract a larger carrier density than the one expected from the surface state dispersion as evidenced in ARPES experiments or corresponding to the SdH oscillations. Moreover, Hall data must be fitted by considering, at least, two different types of carriers, $n$-type and $p$-type, in order to explain the non-linear Hall effect as explained below.

\subsection{Shubnikov-de~Haas oscillations}

\begin{center}
\begin{figure}[t!]
\includegraphics[scale=0.34]{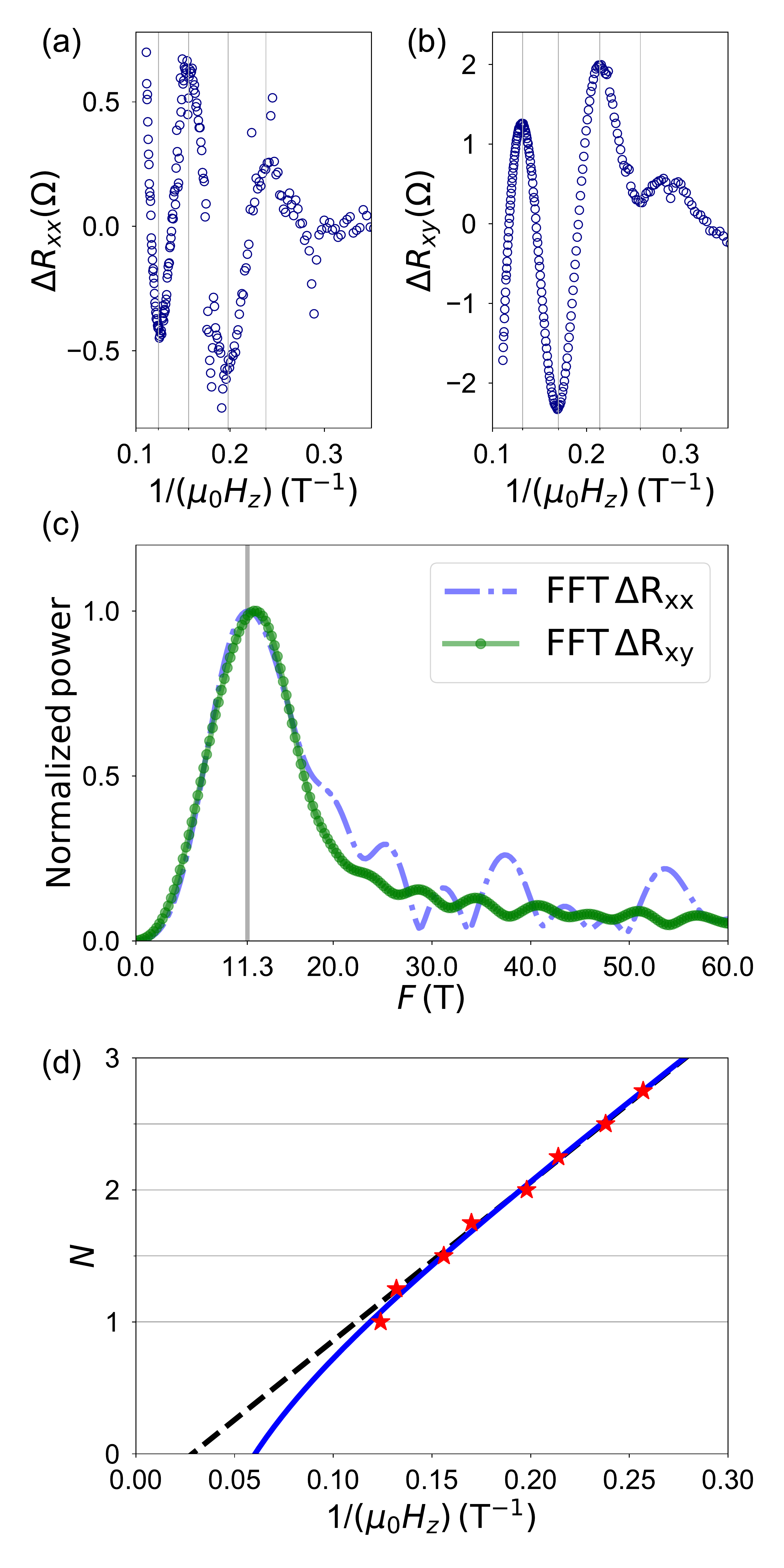}
\caption{Shubnikov-de~Haas oscillations in AlO$_x$ capped $\alpha$-Sn. (a) $\Delta R_{xx}$ and (b) $\Delta R_{xy}$ resistance oscillations as a function of the inverse of the applied perpendicular magnetic field $\mu_0 H_{{z}}$. (c) Fourier transform of both $\Delta R_{xx}$ and $\Delta R_{xy}$ signals displayed in (a) and (b). (d) Landau levels deduced from the peaks in (a) and (b) plotted as a function of $1/(\mu_0 H_{{z}})$ and fitted using the conventional linear relation (black dashed line) and a corrected version using Eq.(\ref{MCKenzieEq}) (blue line) with its linear part (blue doted line) suggesting Dirac fermions.}
\label{figure2}
\end{figure}
\end{center}

Typical SdH oscillations acquired on AlO$_x$ encapsulated samples are displayed in Fig.~\ref{figure2}(a-b). Those plots have been obtained from the acquisition of both sheet and transverse resistances data after subtraction of a polynomial function leading to the determination of two relevant resistance variations $\Delta R_{xx}$ and $\Delta R_{xy}$ (the current $I$ is applied along the $x$ direction whereas the voltage is acquired along the $x$ or $y$ direction depending on the configuration measurement). As expected for SdH oscillations, $R_{xx}$ and $R_{xy}$ are ``phase shifted" from each other by a phase equal to $\pi/4$ as discussed by Wright and McKenzy~\cite{McKenzy13}.

Performing a Fourier transform of the resistance data, one can estimate the characteristic frequency $F$ of the oscillations, in the present case $F = 11.3$\,T [Fig.~\ref{figure2}(c)]. Using the Onsager relationship, one can extract the characteristic cross sectional area of the Fermi surface in the plane of the reciprocal space normal to the magnetic field direction, A$_{\rm F}$, according to the expression~\cite{Onsager52}:
\begin{eqnarray*}
F=\frac{\Phi_{0}}{2\pi^2}A_{\rm F}\quad ,
\end{eqnarray*}
where $\Phi_{0}=h/(2e)$ is the quantum of magnetic flux. Using the circular cross section of the Dirac cone, one can then estimate the characteristic Fermi wave vector to $k_{\rm F}= \sqrt{2e{F}/\hbar}=0.18$\,nm$^{-1}$. On the other hand, considering that the 2-dimensional (2D) carrier density may be expressed as $n_{\rm 2D}=k_{\rm F}^2/(4\pi)$ for the single-band Dirac surface states expected in a TI system, one finds $n_{\rm 2D}=2e{F}/(\hbar\,4\pi) \approx (2.7\pm0.3)\,10^{11}$\,cm$^{-2}$. Using the value of $k_{\rm F}$ derived from the SdH oscillations, in the case of a linear dispersion, one can estimate, the band filling and consequently the Fermi level at $E_{\rm F}^{\rm SdH}=\hbar v_{\rm F} k_{\rm F} = 63$\,meV. This value is in close agreement with the respective values, $E_{\rm F}^{\rm ARPES}=75 \pm 25$\,meV and $k_F = (0.22 \pm 0.08)$\,nm$^{-1}$, extracted from ARPES analysis.

Fig.~\ref{figure2}(d) displays the number $N$ of occupied Landau levels \textit{vs.} the inverse of the magnetic field ($1/(\mu_0 H_z)$). Here, integer values correspond to minima in the oscillations of $\Delta R_{xx}$ [Fig.~\ref{figure2}(a)] whereas half-integers values correspond to maxima; this is due to the bulk band contribution to the resistivity~\cite{Xiong12}. Due to the ``phase shift" between $R_{xy}$ and $R_{xx}$ oscillations, we have indexed the resistance oscillations extrema of $\Delta R_{xy}$ with respect to the Landau level positions shifted by $\pm1/4$ [Fig.~\ref{figure2}(b)]~\cite{McKenzy13} as discussed above. The expression of the Shubnikov-de~Haas oscillations of the longitudinal resistance corresponds to:
\begin{eqnarray*}
\Delta R_{xx} \propto \cos\left[ 2 \pi \left( \frac{F}{\mu_0 H_{{z}}} - \gamma \right) \right]\quad ,
\end{eqnarray*}
where $\gamma$ is a phase offset related to the Berry's phase acquired by the carriers during a cyclotron orbit~\cite{McKenzy13}. In the case of normal fermions, this phase offset is strictly $1/2$, whereas in the case of electron-hole symmetric Dirac fermion dispersion $\gamma=0$~\cite{McKenzy13}. Usually, $\gamma$ is deduced from the intercept of the linear fit of the Landau levels positions, $N$, with respect to $1/(\mu_{0} H_z)$. From our measurements a linear fit leads to $\gamma\simeq -0.3$ [dashed line Fig.~\ref{figure2}(d)]. As proposed recently by Wright and McKenzie~\cite{McKenzy13}, in the case of a 3D-TI with broken electron-hole symmetry, a deviation from the linear dependence is indeed expected. A more accurate fit was hence proposed :
\begin{eqnarray}
N=\frac{F}{\mu_0 H_{{z}}} + A_1 + A_2 \mu_0 H_{{z}}\quad .
\label{MCKenzieEq}
\end{eqnarray} 
The topologically relevant phase offset is $A_1$. The condition $A_1=0$ would indicate that the surface states contain a Dirac component with a Berry's phase of $\pi$. The best resulting fit using Eq.~(\ref{MCKenzieEq}) is displayed on Fig.~\ref{figure2}(d) and leads to the determination of $A_1 = 0.011$ and $A_2 = -0.041$ whereas the frequency $F=11.3$\,T is fixed by the Fourier transform. The asymptotic low field limit is equivalent to $A_2 \rightarrow 0$; this yields a straight line [blue dotted line in Fig.~\ref{figure2}(d)].
From $A_1 \simeq 0$, we can thus conclude that the SdH oscillations reveal the Dirac fermions nature of the resistance oscillations with a characteristic Berry's phase of $\pi$.

As a partial conclusion, the analysis of SdH oscillations, assuming that they originate from 2D surface\footnote{We performed the angular dependence of the SdH oscillations. Unfortunately, for angles larger than $15^{\circ}$ from the normal of the film, the oscillations amplitude are damped and it is too difficult to reliably analyze them.}, corroborates the ARPES measurements performed on the $\alpha$-Sn surface coated with AlO$_x$: we consistently deduce the same carrier density of about $n=3 \, 10^{11} \rm cm^{-2}$ and a Berry's phase corresponding to Dirac fermions.

\subsection{Hall and magnetoresistance}

\begin{center}
\begin{figure}[htb]
\includegraphics[scale=0.34]{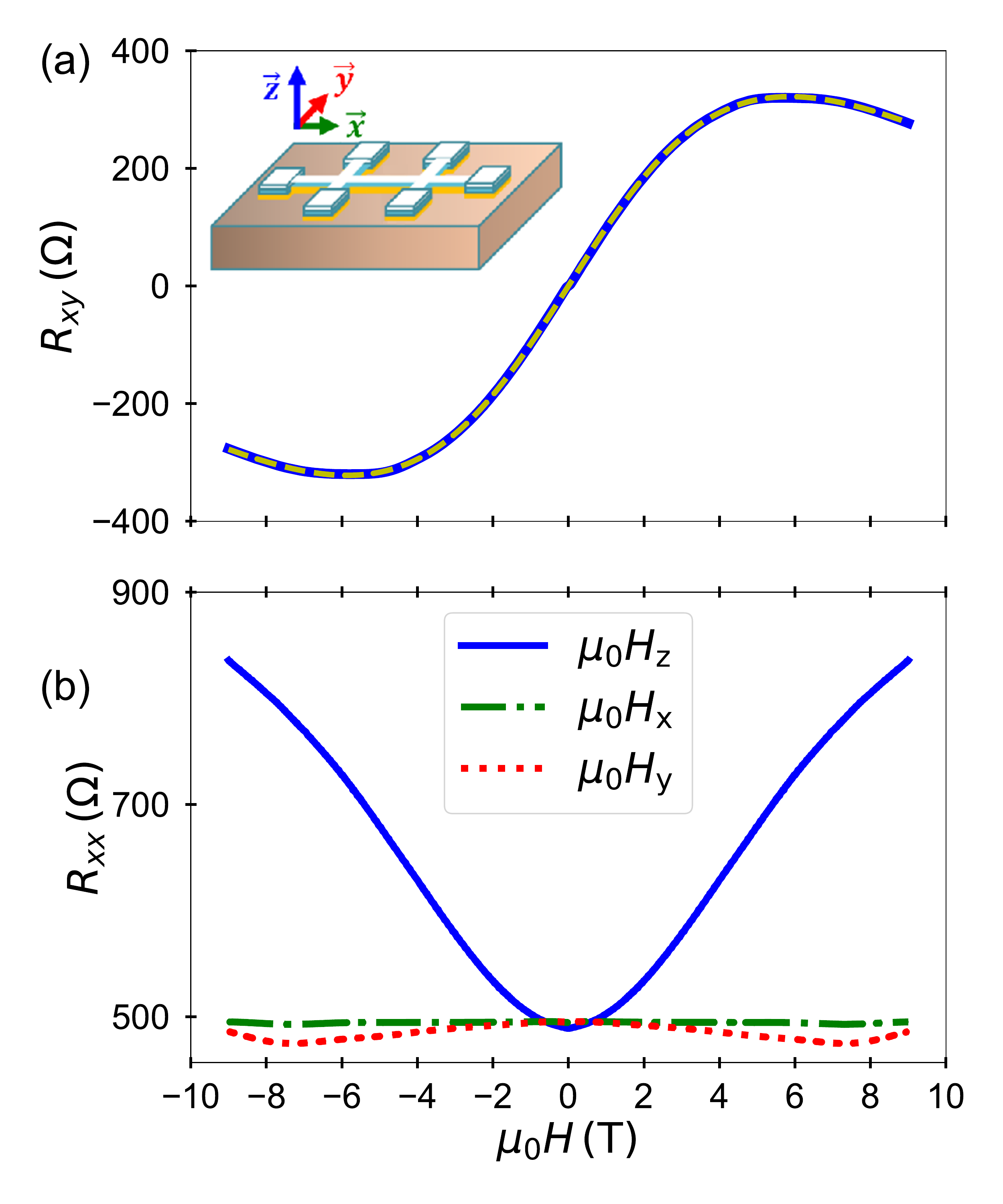}
\caption{Longitudinal and transverse resistance measurements of the
  Hall cross bar.(a) Transverse resistance $R_{xy}$ (Hall) measurement of the
  Hall cross using a perpendicular magnetic field ($H_z$) with the
  corresponding fit using Eq.\eqref{RxyFormula} for a 3-band model
  in dashed yellow line. A sketch of the device configuration is displayed in inset. (b) Sheet resistance $R_{xx}$ measurement of the Hall bar as a function of the field along the directions illustrated in inset of panel (a).}
\label{figure3}
\end{figure}
\end{center}

We discuss now, the Hall and magnetoresistance data acquired on the same AlO$_x$ encapsulated $\alpha$-Sn sample used in previous experiments (ARPES and SdH).  Hall measurements displayed in Fig.~\ref{figure3}(a) indicate the presence of more than one conduction band. The typical change of slope at ``low'' and ``high'' fields is often modeled by two parallel conduction bands with different carrier concentrations and mobilities. The high mobility$/$low carrier density would be ascribed to the states detected by SdH. We are now going to consider first a 2-band and then a 3-band model to fit with both conduction and magneto-transport data.

The first strategy is to fit the Hall data using a simplified two bands model according to the following formula giving the respective longitudinal, $G_{xx}$, and transverse, $G_{xy}$, conductivities:
\begin{eqnarray}
\label{RxyFormula}
\nonumber {G}_{xx} &=&\dfrac{n_1 e \mu_1}{1+(\mu_{1} \mu_0
  H_{{z}})^{2}}+\dfrac{n_2 e \mu_2}{1+(\mu_{2} \mu_0 H_{{z}})^{2}}
\\\nonumber {G}_{xy} &=&\dfrac{n_1 e \mu_1^2 \mu_0 H_{{z}}}{1+(\mu_{1}
  \mu_0 H_{{z}})^{2}}+\dfrac{n_2 e \mu_2^2 \mu_0 H_{{z}}}{1+(\mu_{2}
  \mu_0 H_{{z}})^{2}} \\ R_{xy} &=& \dfrac{{G}_{xy}}{{G}_{xy}^2 +
  {G}_{xx}^2} \hspace{0.8cm} R_{xx} = \dfrac{{G}_{xx}}{{G}_{xy}^2 +
  {G}_{xx}^2 }
\end{eqnarray}
with the constraint of the known sheet resistance at zero field $R_s (H=0) = [\Sigma_i (n_i e \mu_i)]^{-1}$. 

Fitting the data with this procedure results in bands of $n$- and $p$-type with carrier concentration and mobilities of $n_1 = 2.3\,10 ^{12}$\,cm$^{-2}$, $\mu_1 = 3730$\,cm$^2$/(Vs) and $n_2 = 5\,10^{13}$\,cm$^{-2}$, $\mu_2 = 110$\,cm$^2$/(Vs) respectively.
However, the result of the fit is not satisfactory because the carrier density extracted from the $p$-type band is not in agreement with the one extracted from SdH data. This indicates that the proposed two bands model is not sufficient to describe completely the physical processes in our samples. We also note that SdH oscillations can only be observed if full orbits are possible in real space, {\it i.e.}, for mobilities larger than about $2500$\,cm$^2$/(Vs) for oscillations starting at $4$\,T.

We henceforth consider a 3-band model for magnetotransport in order to account for the surface states conduction. Fixing one $p$-type band: $n_{\rm SdH} = (2.7 \pm 0.3)\,10^{11}\,{\rm cm}^{-2}$ corresponding to the SdH data, the best fit, displayed in Fig.~\ref{figure2}(a), gives a corresponding mobility $\mu_{\rm SdH} =2540$\,cm$^2$/(Vs), satisfying now the conditions to observe the SdH oscillations above about 4\,T. Because of the huge difference in the carrier density, $n_{\rm SdH} \ll n_2$ still of $p$-type, the resulting fit of $R_{xy}(H_z)$ does not change much the parameters of the two other bands. ARPES [Fig.~\ref{figure1}.(a) and (b)] and calculations by {\it ab-initio} techniques in $\alpha$-Sn (as discussed in section IV), are consistent with the presence of surface states in parallel to bulk (3D) bands due to $\alpha$-Sn ($p$-type). The $n$-type band has been ascribe to the InSb layer. Indeed, from prior Hall transport measurements performed on an InSb(100nm)$|$GaAs ``substrate" without any specific surface preparation, one could deduce an $n$-type conduction with a carrier density of $n=1.4\,10^{12}$\,cm$^{-2} \, \simeq n_1$ with a mobility of $\mu=3180$\,\muu{} $\simeq \mu_1$.

It becomes now possible to estimate the sheet resistance of the surface states $R_{\rm SdH}=1/(n_{\rm SdH} \, e \, \mu_{\rm SdH}) = 9.2$\,k$\Omega$. This contribution to the resistance should be compared to the contribution from the two other bands and deduced from the three bands fit (carrier densities and mobilities); given by $R_{n-\rm{type}} = 766\,\Omega$ (substrate contribution) and $R_{p-\rm{type}} = 1.3$\,k$\Omega$ (bulk state contribution). From the sheet resistance of the surface states, and using the simple expression of the conductance given for non-degenerate 2D materials with a linear dispersion, one finds:
\begin{eqnarray*}
\sigma_{\rm TI} = \dfrac{1}{R_{SdH}} = \dfrac{e^2}{4 \hbar ^2} \dfrac{E_F}{\pi}\tau \quad,
\end{eqnarray*}
where $\tau$, the momentum relaxation time being evaluated to:
\begin{equation}
\tau = \dfrac{4}{R_{\rm SdH}}\dfrac{\pi \hbar ^2}{e^2 E_F}\approx 60\,{\rm fs}\quad .
\label{tau}
\end{equation}

This relaxation time is more than one order of magnitude larger than the one measured for $\alpha$-Sn covered with Ag.\cite{RojasSanchez16} It emphasizes the specific feature that a metallic capping layer such as Ag may constitute a parallel momentum relaxation channel. Such parallel relaxation mechanism for the carrier momentum may be prevented when an insulating encapsulating AlO$_x$ layer is used instead.

We now focus on the magnetoresistance in the same AlO$_x$ encapsulated $\alpha$-Sn sample, which is displayed in Fig.~\ref{figure3}(b). We observe a strong positive magnetoresistance of about $80 \%$ at 9T when the field is applied normal to the surface, whereas a transverse in-plane magnetic field gives smaller negative values of about $-10 \%$ at 9T. Importantly, the magnetoresistance of the substrate is measured to be significantly smaller, of the order of 29\% at 9\,T. 

We deduce that the large out-of-plane magnetoresistance can not be only explained using the parameters deduced from the 3-band model suggesting an intrinsic effect from the $\alpha$-Sn.
However, a full understanding of the anisotropic
magnetoresistance is beyond the scope of the present paper.

\section{First-principles calculations}

\begin{figure}[htb]
\begin{center}
\resizebox{8.5cm}{!}{\includegraphics{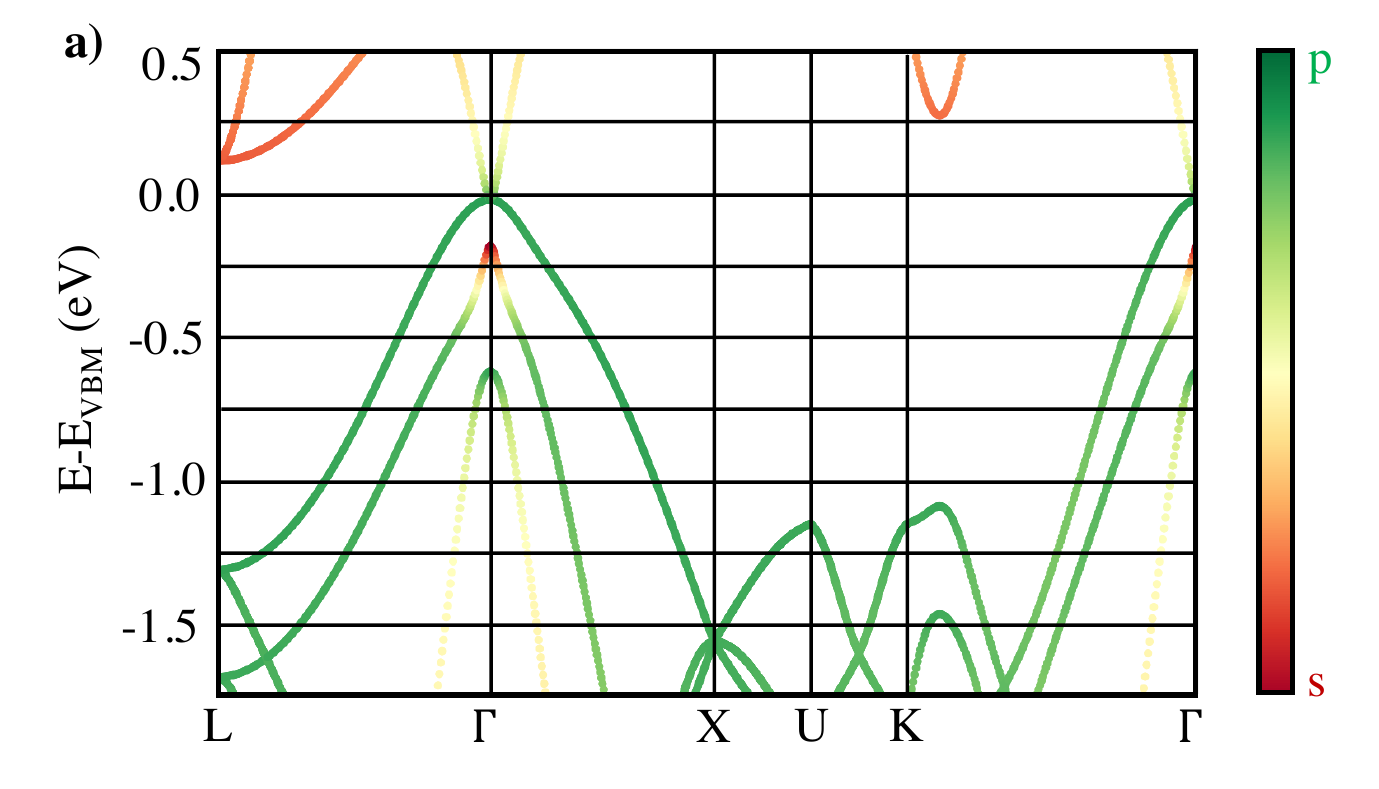}}
\caption{{\it First-principles} band structures of bulk $\alpha$-Sn 
  along the high symmetry points of the Brillouin zone. Bands are
  projected on the $s$ (red) and $p$ (green) orbital characters. }
\label{figDFT1}
\end{center}
\end{figure}

The {\it first-principles} calculations of $\alpha$-Sn electronic properties based on the Density
Functional Theory (DFT) were performed using the VASP package~\cite{VASP1,VASP2}
with the PBE functional revised for solids~\cite{PBEsol} in
combination to an effective $U_{\rm eff}$ potential on Sn-$p$ levels
that we considered equal to $-2$\,eV~\cite{Barfuss13}. The
energy cut-off is set to $400$\,eV. For bulk materials, we used a
$8\times8\times8$ $k$-point mesh and we relaxed the structure until
forces were lower than 0.1\,meV\,\AA$^{-1}$. In order to identify the
topological insulating state of $\alpha$-Sn, we used slabs geometries
consisting of a 12 bulk $\alpha$-Sn unit cells grown along the [0\,0\,1]
direction, {\em i.e.} 48 Sn ML. Surfaces are separated by at least 20
\AA{} of vacuum in order to avoid interaction between the replica of
the slabs. The external slabs are further passivated with two H atoms per Sn
atoms. The $k$-point mesh is downgraded to $8\times8\times1$
points. The geometry relaxation is performed until forces were lower
than 0.01\,eV\,\AA$^{-1}$ and only the outer 12 Sn ML as well
as H positions were allowed to relax (no relaxation of in-plane lattice
parameters were performed). The Sn positions in the inner part of the
slabs were fixed at the calculated bulk parameter values, that we preliminary
optimized with our DFT parameters. Spin-orbit interaction is included
in the simulations. Band structures were plotted with the PyProcar
script~\cite{PyProcar}, using dense $k$-point meshes along the
reciprocal space high symmetry directions.

We first briefly inspect the bulk properties of $\alpha$-Sn. The
geometry relaxation using the PBEsol functional plus a $U$ potential
yields a lattice parameter of 6.4142 \AA, in perfect agreement with
experimental values (0.99\% of error). Furthermore, our DFT
simulations predict that $\alpha$-Sn is a zero gap semiconductor and
all the key features of the band structure are
captured~\cite{PRB-57-1505-1998}: (i) we predict the correct band
order around the $\Gamma$ point, {\em i.e.}, three bands with dominant
$p$, $s$ and $p$ character going from the Fermi level to deeper
energies, as well as, (ii) the absence of electron pockets at the zone
boundary [Fig.~\ref{figDFT1}]. This is a clear improvement over
standard LDA and GGA calculations, thus validating our choice of $U$
potential.

\begin{figure}[htb]
\begin{center}
\resizebox{8.5cm}{!}{\includegraphics{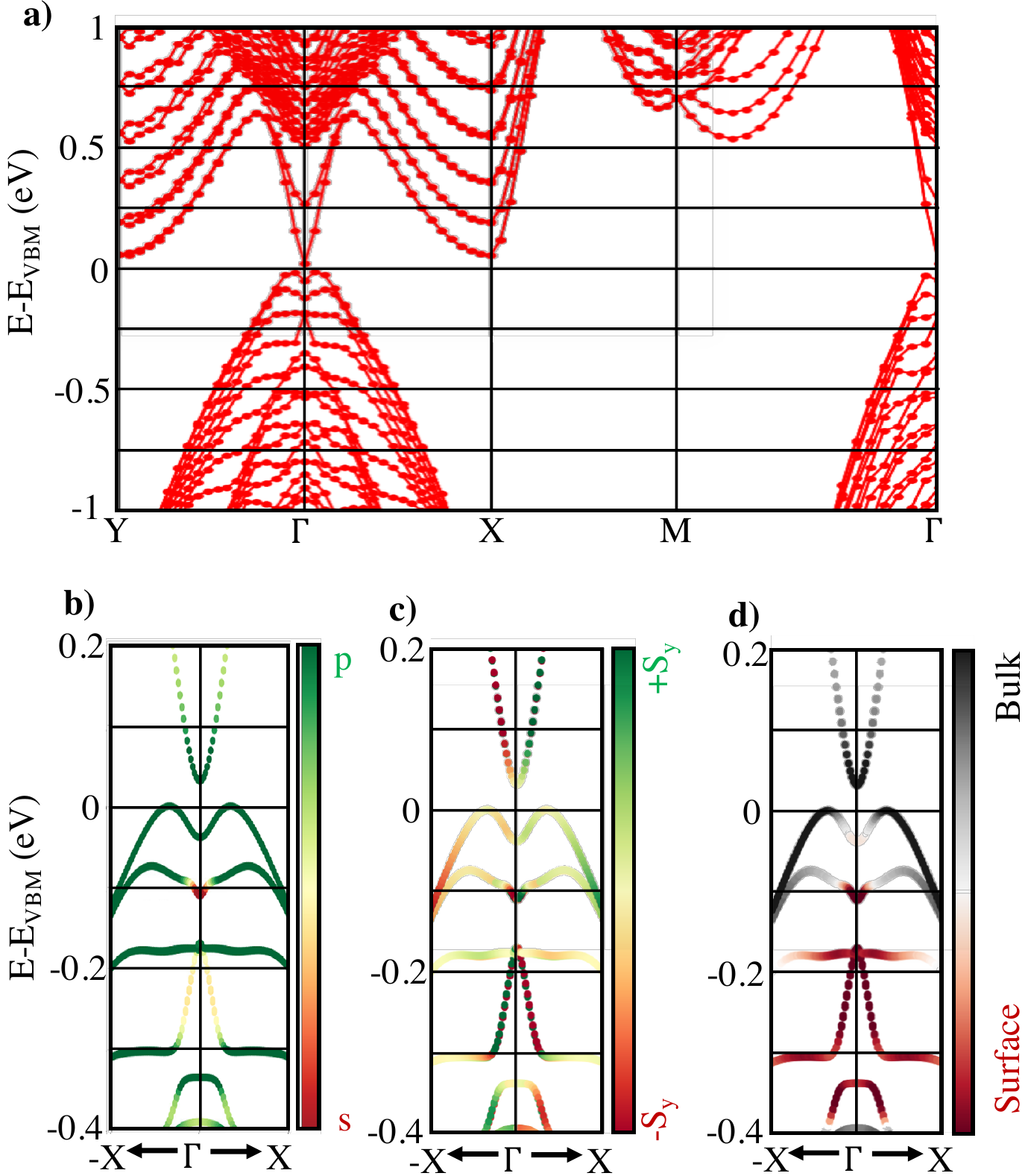}}
\caption{{\it First-principles} band structures of $\alpha$-Sn in a
  48\,ML Sn slab configuration with bulk in-plane lattice
  parameters (a-d). (a) Band structure projected in the reciprocal
  high symmetry directions.  Y=(0,1/2,0), $\Gamma$=(0,0,0),
  X=(1/2,0,0) and M=(1/2,1/2,0). (b) Band structure projected on the
  $s$ (red) and $p$ (green) orbital components along the $-k_{x}$ to $+k_{x}$
  direction. (c) Band structure projected on the $\pm S_y$ spin flavor
  along the $-k_x$ to $+k_x$ direction. (d) Band structure projected
  on the outer 8\,ML Sn atoms, thus forming the surface, along the $-k_x$
  to $+k_x$ direction.}
\label{figDFT2}
\end{center}
\end{figure}

We now explore slabs properties whose in-plane lattice parameter is
fixed to the bulk value, without epitaxial strain. We trace in
Fig.~\ref{figDFT2}(a) the band structure of the material. The slab is
found to be insulating with a narrow band gap of $60$\,meV, in sharp
agreement with earlier DFT studies~\cite{Barfuss13}. We
observe an incipient Dirac state below the top of the valence band,
located between the Fermi level and $-0.25$\,eV. To evidence the presence of the Dirac cone, we projected the band structure along the $-k_x$ to
$+k_x$ direction [Fig.~\ref{figDFT2}(b-d)]. We observe $\pm S_y$ spin
polarized states that likely produce a Dirac cone, that is however ``non closed". This fact is not new and was already found with more sophisticated hybrid functional calculations\cite{PRB-90-125312-2014}. It possibly results from localization errors. In particular we checked that PBEsol calculations without a
$U$ parameter exhibit a closed Dirac cone using the same atomic
positions. The Dirac point is then located roughly $130$\,meV below the
Fermi level and most notably, it is buried inside bulk bands
[Fig.~\ref{figDFT2}(d)]. Secondly, we inspected the role of the
compressive strain ($\varepsilon_{xx}=\varepsilon_{yy}=-0.14$\%)
induced by the InSb substrate on the band structure of the
material. We do not observe any significant changes of the electronic
band order, although the gap increases to $93$\,meV. Finally, we
studied the role of the capping layer by substituting the H atoms by O
atoms (thus forming a SnO$_2$ layer at the surface). Again, no substantial
modifications are raised by these additional simulations (with the finding of a gap value close to $64$\,meV).

DFT simulations thus predict that 48\,ML of Sn form a
topological insulator. This is in good agreement with the ARPES
and transport measurements on $\alpha$-Sn films covered with
AlO$_x$. The shift of the Fermi level 
[Fig.~\ref{figure1}(b)] observed experimentally introduces
additional bands of $p$-type carriers [Fig.~\ref{figure3}(a)]
characterized by a bulk density of $\simeq 10^{13} \rm{cm}^{-2}$. As
evidenced in Fig.~\ref{figDFT2}(d), those additional bulk bands
appears below the Dirac point when increasing the $k$ vector. It could explain the origin of the $p$-type conduction. For a slab calculation using relaxed in-plane lattice parameters, by moving the Fermi energy at $-75$\,meV below the Dirac point in the theoretical investigations, we found a carrier density of $\simeq 5 \, 10^{12}$\,cm$^{-2}$, a value of the same order of the experimental value. It suggests that these bands have about the same mobility and hence can be fit with only one $p$-type bulk band in the Hall effect measurements.

\section{Conclusion}

Using ARPES measurements, we demonstrate that the Fermi level position, relative to the DP, changes depending on the capping layer. In the present case, the drawback for the conduction measurements of capping with AlO$_x$ is that, with a Fermi level below the DC, bulk bands now coexist with the topological surface states. Our {\it ab-inito} calculations confirm the presence of Dirac cone and the existence of a $p$-type bulk band below the Dirac point. It is in agreement with Hall measurements where such a $p$-type band was identified. From transport measurements and SdH oscillations, we could evidence the signature of topological surface states. It is then possible to estimate the relaxation time of the surface states: $\tau=60$\,fs. Compared with the previous measurements for $\alpha$-Sn covered by Ag layer~\cite{RojasSanchez16}, $\tau$ is more than 10 times larger when
$\alpha$-Sn is capped with AlO$_x$. As $\tau$ is still definitely
shorter than that derived from ARPES for the free
surface~\cite{Hajlaoui2014}, we can speculate that the existence of
bulk state at the Fermi level contributes to its shortening by
inter-band scattering. The lifetime is still one order of magnitude larger in the present case compared to Ag capping, because the density of states in the bulk Sn is much lower than in a good metal. Finding other interfacial materials which would keep
the Fermi level above the DP while maintaining the relaxation rate of
the topological surface states at the level of the free surface is an important challenge for
spintronic applications.

\section{acknowledgement}

The authors acknowledge fruitful discussions with G. Bihlmayer, B. Assaf, D. Perconte and P. No\"{e}l. This work was supported by the
French {\it Agence Nationale de la Recherche} through project ANR-16-CE24-0017 {\it TOP-RISE} and the French RENATECH network.

\appendix*
\section{}
\begin{figure}[htb!]
\begin{center}
\resizebox{8cm}{!}{\includegraphics{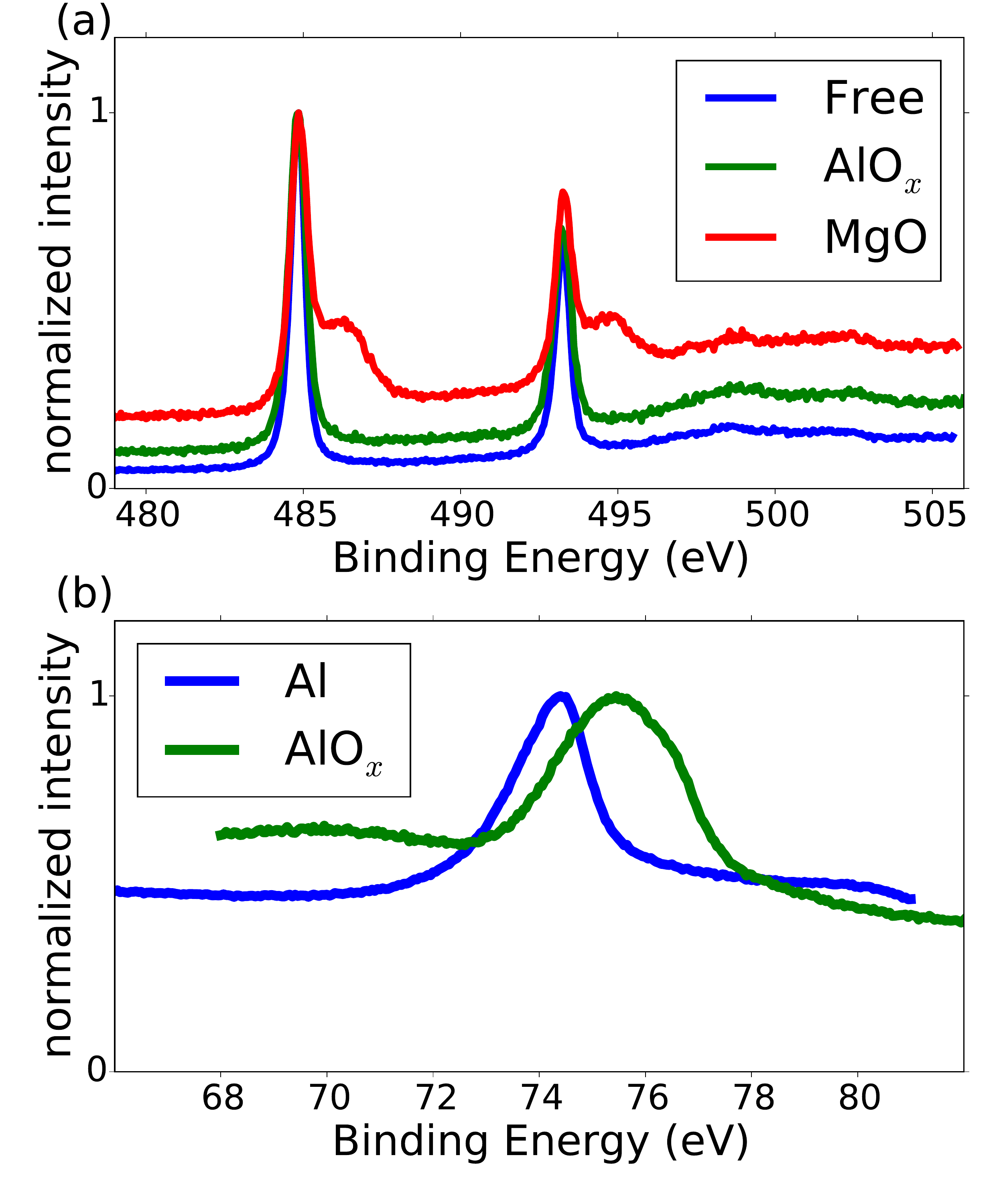}}
\caption{Core level spectroscopy of (a) Sn 3$d_{3/2}$ and  3$d_{5/2}$ peaks before and after deposition of MgO or AlO$_x$; and (b) Al 2$p$ peak before and after exposure to O$_2$ for 40 minutes at $10^{-6}$\,mbar.}
\label{XPS}
\end{center}
\end{figure}
The core level spectroscopy allows one to assess the chemical environment of the probed atoms. In particular, oxidation results in shifts in the binding energy. The Sn 3$d_{3/2}$ and 3$d_{5/2}$ core level spectra for free $\alpha$-Sn surface and when covered either by AlO$_x$ or by MgO is displayed in Fig.~\ref{XPS}(a). On the free surface, both lines are symmetric whereas with an MgO capping, a high binding energy structure appears indicating a partial oxidation of the Sn layer, located at the Sn$\vert$MgO interface. An AlO$_x$ capping leaves both lines unchanged showing that the $\alpha$-Sn layer was protected from oxidation by the Al coating.

The AlO$_x$ coating was prepared by first depositing a 1\,ML metallic Al layer which was subsequently oxidized by a 40\,minutes exposure to molecular O$_2$ at a pressure of $10^{-6}$\,mbar.

The Al 2$p$ levels displayed in Fig.~\ref{XPS}(b) demonstrate the oxidation of the 1\,ML Al film: before oxidization Al 2$p$ peaks around 74\,eV (binding energy), and oxidation leads to a shift of the peak position around 76\,eV as expected~\cite{Rodel}.

%\bibliography{bibli}


\end{document}